\documentclass[ twocolumn, tighten, times, astrosymb]{aastex631}

\usepackage[utf8]{inputenc}

\newcommand{\HST}{$HST$}
\newcommand{\eazy}{{\tt EAzY}}

\newcommand{\OVI}{[\hbox{{\rm O}\kern 0.1em{\sc vi}}]}
\newcommand{\Lalpha}{Ly-$\alpha$}
\newcommand{\NV}{\hbox{{\rm N}\kern 0.1em{\sc v}}}
\newcommand{\SiIV}{\hbox{{\rm Si}\kern 0.1em{\sc iv}}}
\newcommand{\OIV}{[\hbox{{\rm O}\kern 0.1em{\sc iv}}]}
\newcommand{\NIV}{[\hbox{{\rm N}\kern 0.1em{\sc iv}}]}
\newcommand{\CIV}{\hbox{{\rm C}\kern 0.1em{\sc iv}}}
\newcommand{\HeII}{\hbox{{\rm He}\kern 0.1em{\sc ii}\kern 0.1em{$\lambda1640$} }}
\newcommand{\OIII}{[\hbox{{\rm O}\kern 0.1em{\sc iii}}]{$\lambda5007$}}
\newcommand{\NIII}{[\hbox{{\rm N}\kern 0.1em{\sc iii}}]}
\newcommand{\AlIII}{\hbox{{\rm Al}\kern 0.1em{\sc iii}}}
\newcommand{\SiIII}{\hbox{{\rm Si}\kern 0.1em{\sc iii}}}
\newcommand{\CIII}{\hbox{{\rm C}\kern 0.1em{\sc iii}]}}
\newcommand{\NeIV}{[\hbox{{\rm Ne}\kern 0.1em{\sc iv}}]}
\newcommand{\MgII}{\hbox{{\rm Mg}\kern 0.1em{\sc ii}}}

\newcommand{\CII}{[\hbox{{\rm C}\kern 0.1em{\sc ii}]}}

\newcommand{\He}{\hbox{{\rm He}\kern 0.1em{\sc ii}\kern 0.1em{$\lambda1640\lambda4686$}}}

\newcommand{\Halpha}{H$\alpha$}

\newcommand{\SII}{[\hbox{{\rm S}\kern 0.1em{\sc ii}}]$\lambda6717\lambda6731$}
\newcommand{\NII}{[\hbox{{\rm N}\kern 0.1em{\sc ii}}]}
\newcommand{\OII}{[\hbox{{\rm O}\kern 0.1em{\sc ii}}]}

\newcommand{\MgI}{\hbox{{\rm Mg}\kern 0.1em{\sc i}}}
\newcommand{\FeII}{\hbox{{\rm Fe}\kern 0.1em{\sc ii}}}

\newcommand{\OI}{\hbox{{\rm O}\kern 0.1em{\sc i}}}
\newcommand{\NeII}{[\hbox{{\rm Ne}\kern 0.1em{\sc ii}}] }
\newcommand{\NaI}{[\hbox{{\rm Na}\kern 0.1em{\sc i}}] }
\newcommand{\NeIII}{[\hbox{{\rm Ne}\kern 0.1em{\sc iii}}] }

\submitjournal{ApJL}

\shorttitle{Discovering a bluer $z\sim4-7$ Universe through UV slopes}
\shortauthors{Nanayakkara et al.}
\graphicspath{{./}{figures/}}

\begin{document}

\title{Early results from GLASS-JWST XVI: Discovering a bluer $z\sim4-7$ Universe through UV slopes}

\author[0000-0003-2804-0648 ]{Themiya Nanayakkara}
\affiliation{Centre for Astrophysics and Supercomputing, Swinburne University of Technology, PO Box 218, Hawthorn, VIC 3122, Australia}

\author[0000-0002-3254-9044]{Karl Glazebrook}
\affiliation{Centre for Astrophysics and Supercomputing, Swinburne University of Technology, PO Box 218, Hawthorn, VIC 3122, Australia}

\author[0000-0003-4239-4055]{Colin Jacobs}
\affiliation{Centre for Astrophysics and Supercomputing, Swinburne University of Technology, PO Box 218, Hawthorn, VIC 3122, Australia}

\author{Andrea Bonchi}
\affiliation{Space Science Data Center, Italian Space Agency, via del Politecnico, 00133, Roma, Italy}

\author[0000-0001-9875-8263]{Marco Castellano}
\affiliation{INAF - Osservatorio Astronomico di Roma, via di Frascati 33, 00078 Monte Porzio Catone, Italy}

\author[0000-0003-3820-2823]{Adriano Fontana}
\affiliation{INAF - Osservatorio Astronomico di Roma, via di Frascati 33, 00078 Monte Porzio Catone, Italy}

\author[0000-0002-3407-1785]{Charlotte Mason}
\affiliation{Cosmic Dawn Center (DAWN)}
\affiliation{Niels Bohr Institute, University of Copenhagen, Jagtvej 128, 2200 København N, Denmark}

\author[0000-0001-6870-8900]{Emiliano Merlin}
\affiliation{INAF - Osservatorio Astronomico di Roma, via di Frascati 33, 00078 Monte Porzio Catone, Italy}

\author[0000-0002-8512-1404]{Takahiro Morishita}
\affiliation{IPAC, California Institute of Technology, MC 314-6, 1200 E. California Boulevard, Pasadena, CA 91125, USA}

\author[0000-0002-7409-8114]{Diego Paris}
\affiliation{INAF - Osservatorio Astronomico di Roma, via di Frascati 33, 00078 Monte Porzio Catone, Italy}

\author[0000-0001-9391-305X]{Michele Trenti}
\affiliation{School of Physics, University of Melbourne, Parkville 3010, VIC, Australia}
\affiliation{ARC Centre of Excellence for All Sky Astrophysics in 3 Dimensions (ASTRO 3D), Australia}

\author[0000-0002-8460-0390]{Tommaso Treu}
\affiliation{Department of Physics and Astronomy, University of California, Los Angeles, 430 Portola Plaza, Los Angeles, CA 90095, USA}

\author[0000-0003-2536-1614]{Antonello Calabr\`o}
\affiliation{INAF Osservatorio Astronomico di Roma, Via di Frascati 33, 00078 Monte Porzio Catone, Rome, Italy}

\author[0000-0003-4109-304X]{Kristan Boyett}
\affiliation{School of Physics, University of Melbourne, Parkville 3010, VIC, Australia}
\affiliation{ARC Centre of Excellence for All Sky Astrophysics in 3 Dimensions (ASTRO 3D), Australia}

\author[0000-0001-5984-0395]{Marusa Bradac}
\affiliation{
University of Ljubljana, Department of Mathematics and Physics, Jadranska ulica 19, SI-1000 Ljubljana, Slovenia}
\affiliation{
Department of Physics and Astronomy, University of California Davis, 1 Shields Avenue, Davis, CA 95616, USA}

\author[0000-0003-4570-3159]{Nicha Leethochawalit}
\affiliation{School of Physics, University of Melbourne, Parkville 3010, VIC, Australia}
\affiliation{ARC Centre of Excellence for All Sky Astrophysics in 3 Dimensions (ASTRO 3D), Australia}
\affiliation{National Astronomical Research Institute of Thailand (NARIT), Mae Rim, Chiang Mai, 50180, Thailand}

\author[0000-0001-9002-3502]{Danilo Marchesini}
\affiliation{
Department of Physics and Astronomy, Tufts University, 574 Boston Ave., Medford, MA 02155, USA}

\author[0000-0002-9334-8705]{Paola Santini}
\affiliation{INAF Osservatorio Astronomico di Roma, Via di Frascati 33, 00078 Monte Porzio Catone, Rome, Italy}

\author[0000-0002-6338-7295]{Victoria Strait}
\affiliation{Cosmic Dawn Center (DAWN), Denmark}
\affiliation{Niels Bohr Institute, University of Copenhagen, Jagtvej 128, DK-2200 Copenhagen N, Denmark}

\author[0000-0002-5057-135X]{Eros Vanzella}
\affiliation{INAF -- OAS, Osservatorio di Astrofisica e Scienza dello Spazio di Bologna, via Gobetti 93/3, I-40129 Bologna, Italy}

\author[0000-0003-0980-1499]{Benedetta Vulcani}
\affiliation{INAF Osservatorio Astronomico di Padova, vicolo dell'Osservatorio 5, 35122 Padova, Italy}

\author[0000-0002-9373-3865]{Xin Wang}
\affil{Infrared Processing and Analysis Center, Caltech, 1200 E. California Blvd., Pasadena, CA 91125, USA}

\author[0000-0002-8434-880X]{Lilian Yang}
\affiliation{Kavli Institute for the Physics and Mathematics of the Universe, The University of Tokyo, Kashiwa, Japan 277-8583}

\begin{abstract}
We use the GLASS-JWST Early Release Science NIRCam parallel observations to provide a first view of the UV continuum properties of NIRCam/$F444W$  selected galaxies at $4<z<7$.
By combining multiwavelength NIRCam observations, we constrain the UV continuum slope for a sample of 401 galaxies with stringent quality controls.  
We find that $>99\%$ of the galaxies are blue star-forming galaxies with very low levels of dust ($Av_{\beta}\sim0.01\pm0.33$).
We find no statistically significant correlation for UV slope with redshift or UV magnitude. 
However, we find that in general galaxies at higher redshifts and fainter UV magnitudes have steeper UV slopes. 
We find a statistically significant correlation for UV slope with stellar mass, with galaxies with higher stellar mass showing shallower UV slopes. 
Individual fits to some of our galaxies reach the bluest UV slopes of $\beta\sim-3.1$ allowed by stellar population models used in this analysis. 
Therefore, it is likely that stellar population models with higher amount of Lyman continuum leakage, AGN effects, and/or Population III contributions are required to accurately reproduce the rest-UV and optical properties of some of our bluest galaxies.  
This dust-free early view confirms that our current cosmological understanding of gradual mass + dust buildup of galaxies with cosmic time is largely accurate to describe the $\sim0.7-1.5$ Gyr age window of the Universe.
The abundance of a large population of UV faint dust-poor systems may point to a dominance of low-mass galaxies at $z>6$ playing a vital role in cosmic reionization. 
\end{abstract}


\section{Introduction} \label{sec:intro}

In the current cosmological picture we expect the first galaxies to emerge $\sim200-300$ Myrs after the Big Bang \citep[e.g.][]{Bromm2011}. 
These galaxies would rapidly buildup their stellar masses driving the chemical evolutionary processes of the Universe \citep[e.g.][]{Madau2014,Dayal2018a}. 
Dust/metals produced as end products of stellar evolution in the early galaxies provide key constraints to how cosmic star-formation would have evolved in the first $\sim3$ billion years of the Universe \citep[e.g.][]{Stark2016}. 
Thus, establishing a relationship between dust and luminosity in the early Universe from UV bright to faint galaxies would provide tight constraints to galaxy evolution and cosmology  \citep[e.g][]{Naidu2019,Bouwens2021a,Bouwens2022b,Finkelstein2022,LeReste2022,Leethochawalit2022}.

Previous \emph{Hubble Space Telescope (HST)} and \emph{Spitzer} observations have been able to detect galaxies in the early Universe shedding light on early cosmic processes (see \citet{Bradac2020a} for a review, also \citet{Stefanon2021a,Strait2021a,Bouwens2022c,Finkelstein2022}).
These observations have limitations that hinder their diagnostic power in the early Universe. 
For example, \emph{HST} observations probe the rest frame UV which may be biased towards young stellar populations unobscured by dust while \emph{Spitzer} can only provide constraints on emission lines or strong spectral features like Balmer breaks \citep[e.g][]{Stefanon2022a}.

Spectroscopic observations of high-$z$ sources with \Lalpha\ provide additional galaxy constraints on gas and dust geometries \citep[e.g.][]{Oesch2015,Matthee2018b,Jung2019a}, albeit with limitations \citep{Leonova2021, Endsley2022a}. 
Atacama Large Millimeter/submillimeter Array (ALMA) has been successful in detecting large numbers of sources at $z>6$ in dust emission with high-efficiency \citep[e.g.][]{Hashimoto2019b,Bouwens2022a}, but galaxy selections for ALMA followup are strongly biased toward UV bright sources \citep[e.g.][]{Bouwens2022a}.
Thus, even with recent advancements, the formation timescales of galaxies in the early Universe and their UV luminosity, mass, and dust evolution are not well constrained \citep{Dayal2022,Finkelstein2022,Tacchella2022}.

The launch of the \emph{James Webb Space Telescope} (JWST) opened up a new window for exploring galaxy evolution in the early Universe. 
With JWST/NIRCam NIR $5\mu m$ selection ($F444W$), JWST can obtain a diverse sample of galaxies in rest-frame  optical wavelengths \citep[e.g.,][]{Jacobs2023a,Yang2022b} at $z\sim4-7$. 
In this analysis we use JWST NIRCam \citep{Burriesci2005a} parallel observations from the GLASS-JWST survey \citep{Treu2022a} to explore the relationship between UV magnitude and UV slope  for galaxies between $z\sim4-7$.

The rest-UV slope is a common indicator for dust attenuation and has been used to explore the buildup of dust in the $z\sim2-10$ Universe \citep[e.g.][]{Reddy2018a}. 
While the UV slope could also be degenerate with metallicity and star-formation history \citep[SFH, e.g.][]{Bouwens2016b}, at fixed UV luminosity galaxies with redder UV slopes have been shown to be dustier compared to their bluer counterparts \citep{Reddy2018a}. 
Additionally, at lower metallicities the indications are that galaxies prefer steeper attenuation curves \citep{Reddy2018a}. 
This means that sight lines of young blue stars in low metallicity galaxies have more dust. 

While significant evidence for dust has been hinted by recent ALMA detections, the UV slope evolution between $z\sim4-10$ with UV luminosity is still unclear \citep[e.g.][]{Wilkins2016c,Roberts-Borsani2022a}. 
Studies of UV fainter galaxies (up to $m_{UV}\sim-14$) are largely from  \emph{Hubble} frontier fields, boosted by lensing magnifications \citep[e.g.][]{Yang2022a}, but limited to small volumes.  
At  brighter magnitudes ($m_{UV}\gtrsim-18$), comparisons between field and cluster (lensed) sample UV slopes are found to be largely consistent \citep{Bouwens2021a,Bouwens2022b,Yang2022a}. 

With GLASS-JWST rest-frame optical selection we can constrain the UV slopes of UV faint/red galaxies to investigate whether galaxies are faint due to presence of large amounts of intervening dust or whether they are intrinsically faint in UV due to low star-formation. 
While Lyman-break galaxies with redder UV slopes of $\beta \sim-1$ have been spectroscopically confirmed at $z\sim7$ \citep{Smit2018}, we do not have strong constraints of the true abundance of such galaxies. 
The presence of UV faint high-mass galaxies with shallow UV slopes could lead to significant implications to mass buildup of galaxies at $z>4$.

In Section \ref{sec:UV_slope}, we describe our sample selection and UV slope calculation process. 
In Section \ref{sec:dust}, we interpret our observations in terms of UV luminosity and dust buildup in the early Universe. 
Finally, in Section \ref{sec:conclusions} provide a summary of our results and discuss how future surveys would provide tighter constraints to the our cosmological understanding. 
Unless otherwise stated, we assume a \citet{Chabrier2003} Initial Mass Function (IMF) and a cosmology with  H$_{0}= 70$ km/s/Mpc, $\Omega_\Lambda=0.7$ and $\Omega_m= 0.3$. All magnitudes are expressed using the AB system \citep{Oke1983}.

\section{Sample Selection and UV slope measurements} \label{sec:UV_slope}

\begin{figure*}
\includegraphics[scale=0.85]{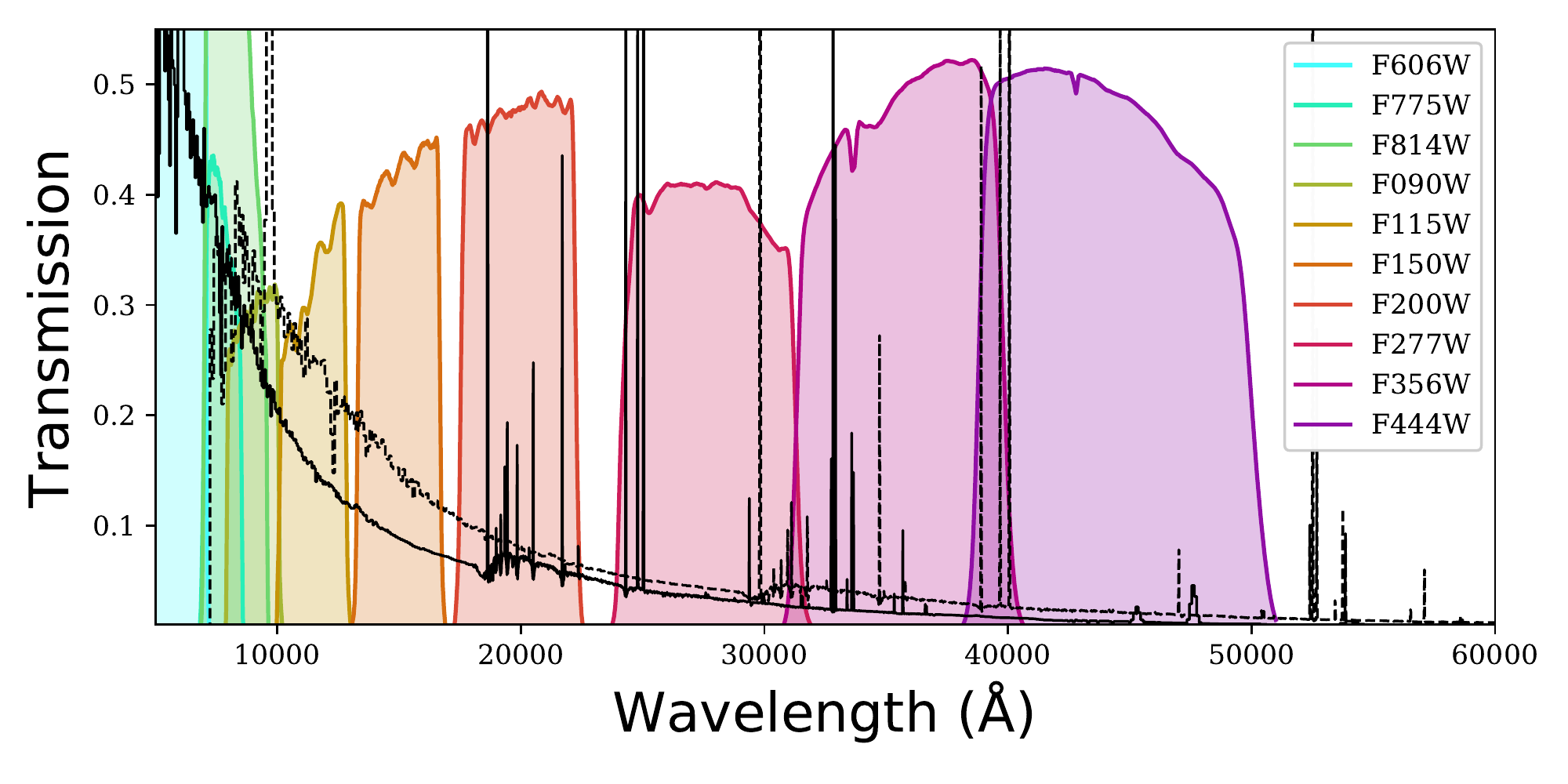}
\includegraphics[scale=1.1]{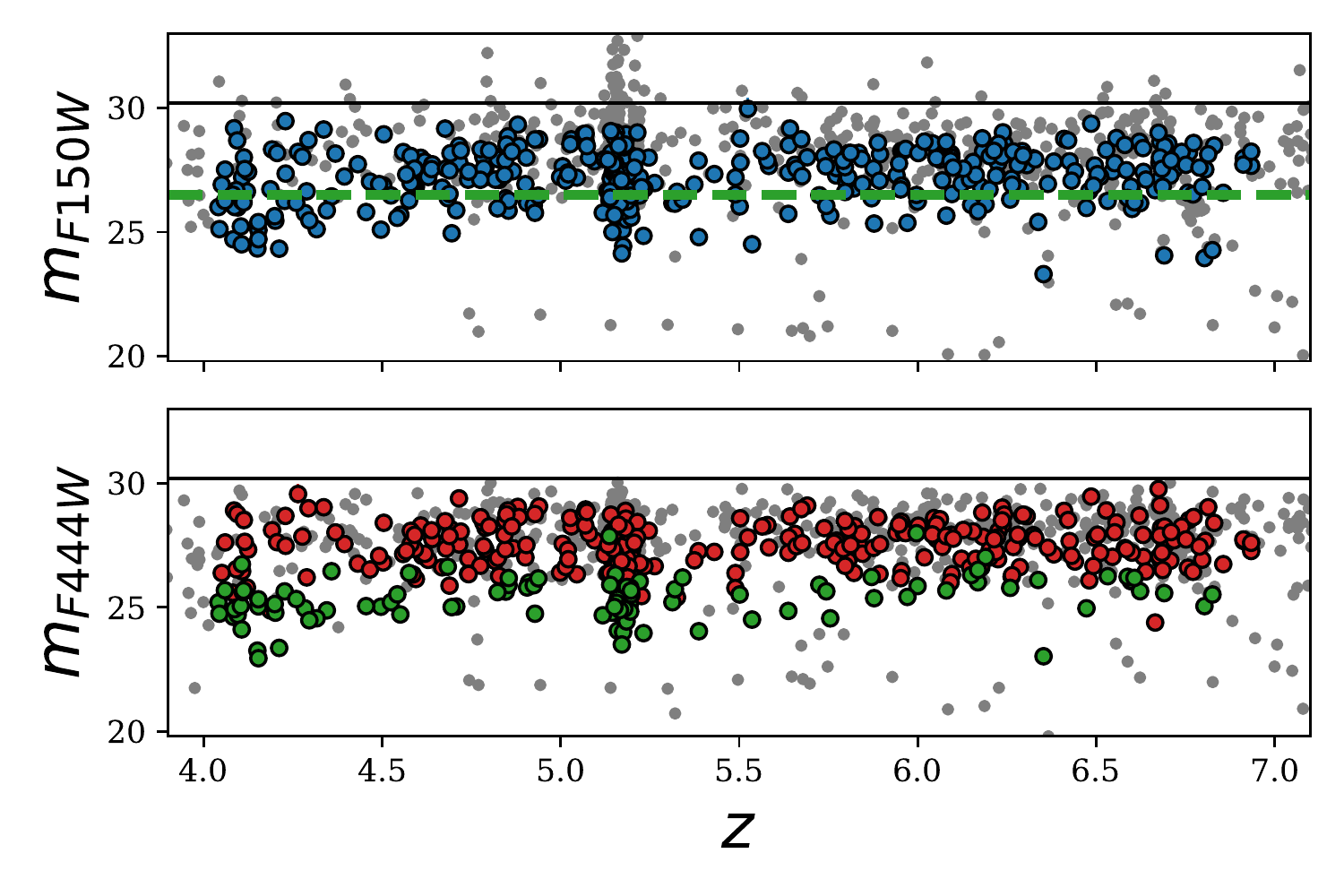}
\caption{{\bf Top:} JWST/NIRCam filter coverage of the GLASS-JWST Early Release Science Program observations. In total, observations are obtained over seven NIR filters reaching a 5-$\sigma$ depth of $m\sim30.2$. We also show an example \eazy\ template of a galaxy at $z=4.0$ (solid lines) and $z=7.0$ (dashed lines), the lower and upper $z$ limits used in our analysis. With this selection, all our galaxies have rest-frame UV and optical coverage which enables stronger constraints to the overall shape of the SED. A majority of the GLASS-JWST NIRCam parallel field is also covered by \emph{HST F606W, F775W,} and \emph{F814W} bands; their filter coverage is also shown here.
{\bf Bottom:} Photometric properties of the sample used in our analysis. We show the magnitude vs redshift distribution of our galaxies in the  ({\bf Bottom Top:}$F150W$) and ({\bf Bottom Lower:}$F444W$) bands observed by our survey. The grey dots are the galaxies removed from our analysis based on the stringent quality cuts outlined in Section \ref{sec:UV_slope}. The horizontal solid line is the nominal 5-$\sigma$ detection level of $m=30.2$ for GLASS-JWST NIRCam parallel field. The horizontal green dashed line in the $F150W$ panel is the $F150W<26.4$ 3DHST survey detection limit \citep{Skelton2014}. We assume that the band pass of HST WFC3/F160W filter is $\sim$ similar to that of the JWST NIRCam/F150W filter. 
Galaxies that were detectable in 3DHST are marked as green circles in the $F444W$ panel. We note that the XDF detection limit of $\sim29.8$ \citep{Bouwens2014a} is similar to that of GLASS-JWST NIRCam parallel field. 
The typical error bars for most galaxies are smaller than the marker size.  
\label{fig:filter_coverage_mag_z}}
\end{figure*}

Our sample is based on the JWST GLASS Early Release Science Program Parallel NIRCam observations \citep{Merlin2022a}  Stage 1 Data Release \citep{Paris2023a}. 
The observations used here were obtained on 28-29th of June and  10–11th of November 2022 using NIRCam filters \emph{F090W, F115W, F150W, F200W, F277W, F356W, F444W}. The data was reduced using the latest NIRCam calibration files ({\tt cjwst 1014.pmap} to {\tt cjwst 1019.pmap}) provided by STScI.
The $F444W$ band was selected for source detection and all images were point-spread function (PSF) matched to this band producing a catalog reaching typical $5$-$\sigma$ depths of $\sim30.2$ mag in all bands.
Additionally, \emph{HST F606W, F775W, F814W, F435W, F105W, F125W, F140W, and F160W} bands and JWST/NIRcam \emph{F410M} bands \citep{Bezanson2022a} were folded in where available over the GLASS-JWST parallel field \citep{Treu2022a}.
More details on the observations, data reduction, and catalog generation can be found in \citet{Paris2023a}.

For the analysis presented in this paper, we required the selected sample to be covered in both rest-frame UV and optical wavelengths in the 7 GLASS-JWST NIRCam bands. 
The rest-frame UV coverage allows us to investigate the accuracy of the best fit spectral energy distribution (SED) shape with the observed photometry.
The rest-frame optical coverage includes the Balmer break and strong emission lines such as \Halpha\ and \OIII.  
Therefore, our final redshift selection for this work is between $4<z<7$. 
Figure \ref{fig:filter_coverage_mag_z} shows the observed band coverage for galaxies in this redshift window along with additional \HST\ data we have obtained over the field \citep{Paris2023a}.

We ran the photometric redshift fitting code \eazy\ \citep{Brammer2008} using the {\tt eazypy} python wrapper\footnote{https://eazy-py.readthedocs.io} to derive photometric redshifts and rest-frame colors for the \citet{Paris2023a} catalogue. 
The total flux measured in 3FWHM PSF matched NIRCam and WFC3 image apertures were selected for this purpose. 
We used the {\tt tweak\_fsps\_QSF\_12\_v3\_}\\{\tt newtemplates\_Lya\_Reduced.param} templates presented in \citet{Larson2022a} with \eazy\ to derive the photometric redshifts and rest-frame colors. These templates are derived using FSPS \citep{Conroy2010a} and BPASS \citep{Eldridge2017} stellar population models and {\tt CLOUDY} \citep{Ferland2017} photoionisation code and are ideal to describe $z\sim4-7$ galaxies\footnote{https://ceers.github.io/LarsonSEDTemplates}. More information on the models can be found in \citet{Larson2022a}.
For each object, we used the maximum likelihood redshift estimated by {\tt eazypy} as the photometric redshift.
Based on 173 secure VLT/MUSE spectroscopic redshifts obtained over the GLASS-JWST field (G. B. Caminha et la., in prep), we derived a photometric redshift accuracy \citep[as defined by][]{Nanayakkara2016,Straatman2016} of $\sim3\%$.
When we include the full \citet{Paris2023a} catalogue (cluster + GLASS-JWST parallel field) with 379 secure spectroscopic redshifts, the photometric redshift accuracy increase to $\lesssim2\%$. 
We also stack the $P(z)$ distributions of $4<z<7$ galaxies used for our analysis and find that there is $\sim87\%$ probability \citep{Nanayakkara2016} for our sample to lie within this redshift window.
We further perform mock GLASS-JWST observations of 5000 galaxies using JAGUAR simulations \citep{Williams2018} and find that the input redshift can be recovered at the $\sim2\%$ accuracy.

GLASS-JWST Stage 1 data release contains 24389 objects \citep{Paris2023a}.
This includes the GLASS-JWST ERS footprint \citep{Treu2022a}, the UNCOVER imaging over the ABELL 2744 cluster \citep{Bezanson2022a}, and additional JWST imaging from JWST DDT 2756 (PI W. Chen).
We select 9272 of galaxies that fall within the GLASS-JWST ERS footprint from the \citet{Paris2023a} catalogue for our analysis, out of which 1024 objects were in the redshift window of interest, $4<z<7$.

We further pruned this sample by applying stringent quality cuts as follows. 
First we required galaxies to be detected with a signal to noise (S/N) $>5$ in at least 4 photometric bands.
This resulted in a sample of 587 galaxies which were detected in a majority of the GLASS-JWST filters. 
We then visually inspected all the \eazy\ fit SEDs of the galaxies to determine that there were no failed fits. 
We defined a fit as a failure if a majority of the observed photometric data points did not agree with the template derived flux within photometric errors, i.e. 1-$\sigma$ outliers. 
Only two galaxies fell into this criteria.
We further investigated the redshift probability distribution of the galaxies and removed sources that had $>50\%$ of the maximum likelihood value at a different redshift. 
This resulted in a sample of 487 galaxies. 
Most galaxies that are removed from the sample lie on the redshift range with the highest number of detections ($z\sim4.5-5.5$). 
Finally, we investigated cutout images of our sample in all the JWST filters used in our analysis and removed sources that were spurious, i.e. sources at the edges of the detectors, sources contaminated with bright neighbors, fragments of stellar spikes, misidentified stars (see \citet{Merlin2022a} for details). 
These stringent cuts resulted in 401 galaxies with high quality photometry and SED fits.

The magnitude distribution of our final sample in the bluest ($F150W$) and reddest ($F444W$) filters as a function of redshift is shown by Figure~\ref{fig:filter_coverage_mag_z}. 
There is a high abundance of galaxies at $z\sim5.1$, but apart from that the redshift distribution is quite flat.
We find that most galaxies removed by our quality cuts are also at $z\sim5.1$. 
Galaxies span a similar range in magnitude in the redder $F444W$ filter ($m_{median}=27.1\pm1.3$) and bluer $F150W$ filters  ($m_{median}=27.4\pm1.1$). 
We also show the $WFC3/F160W=26.4$ 5-$\sigma$ detection level of the 3DHST GOODS-S catalogue \citep{Skelton2014} in Figure~\ref{fig:filter_coverage_mag_z}. GLASS-JWST is several magnitudes deeper compared to 3DHST and reaches similar detection levels of the deepest imaging in the Hubble eXtreme Deep Field (XDF) obtained by \HST\ \citep[\emph{WFC/F160W} 5-$\sigma \sim$29.8][]{Bouwens2014a}.  
The faint galaxies that are not detected by 3DHST $\sim1.5\mu m$  are also faint in the $F444W$ band. 
This suggest that with the GLASS-JWST $F444W$ selection, we are simply obtaining fainter galaxies that would have otherwise been missed by shallower surveys.

\begin{figure*}
\includegraphics[scale=0.6]{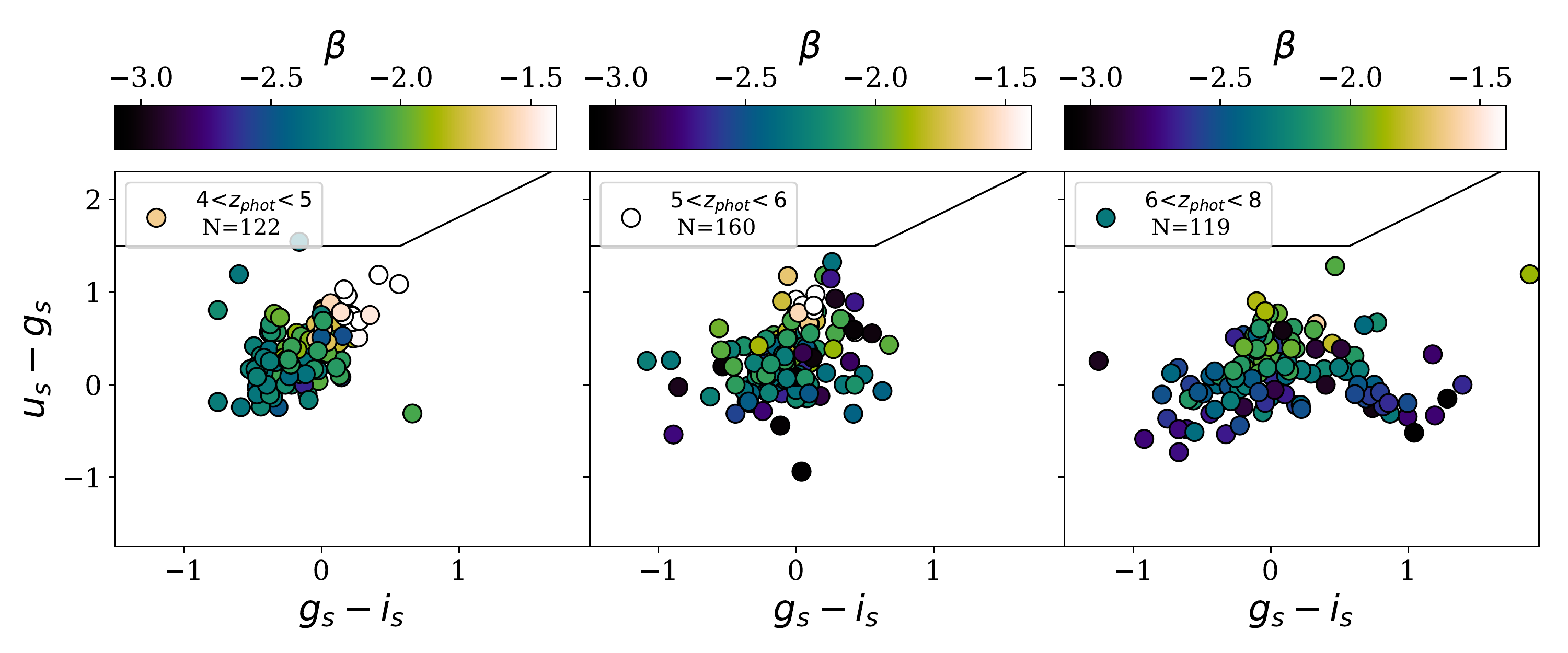}
\caption{
The Rest-frame $u_s-g_s$ vs $g_s-i_s$ ($ugi$) color distribution \citep{Antwi-Danso2022a} of our sample. 
We divide the sample into three redshift bins as annotated in the figures and are color coded based on the UV slope $\beta$.
Almost all galaxies in our sample fall in the star-forming region of the $ugi$ color space. In general galaxies that have bluer $ugi$ colors also show steeper UV slopes. 
\label{fig:colors}}
\end{figure*}

We use the best-fit \eazy\ SEDs to compute the UV slope and UV magnitude of our sample similar to the process outlined by \citet{Nanayakkara2020}. 
We first define a box car filter between $\Delta \lambda= 1400-2000$\AA\ and compute the UV magnitude from the \eazy\ best-fit SEDs. 
Next, for each SED we select a wavelength window between $\Delta \lambda= 1400-2000$\AA\ and mask out emission line regions within this window following masks outlined by \citet{Calzetti1994}.  
We then use the masked SEDs to compute the UV continuum slope $\beta$ by fitting a power-law function using {\tt lmfit} \citep{Newville2014a}. 
We found a single power law to be an accurate description to the UV slopes of all the galaxies in our sample.

We use the JAGUAR mock galaxy catalogue \citep{Williams2018} to investigate the robustness of our UV slope measurements. We select 50,000 galaxies randomly from the JAGUAR catalogue that satisfy the GLASS-JWST $F444W$ selection. 
We then perform  quality controls to the mock data similar to what we did for the real data in the same filters. 
The final sample contains 307 galaxies. 
The intrinsic UV slope $\beta$ reported by \citet{Williams2018} is recovered with a median offset of only $0.02\pm0.11$. 
Thus, our process can recover the UV slopes of galaxies at high accuracy.

\begin{figure*}
\includegraphics[scale=0.45, trim={350 50 5 0}, clip]{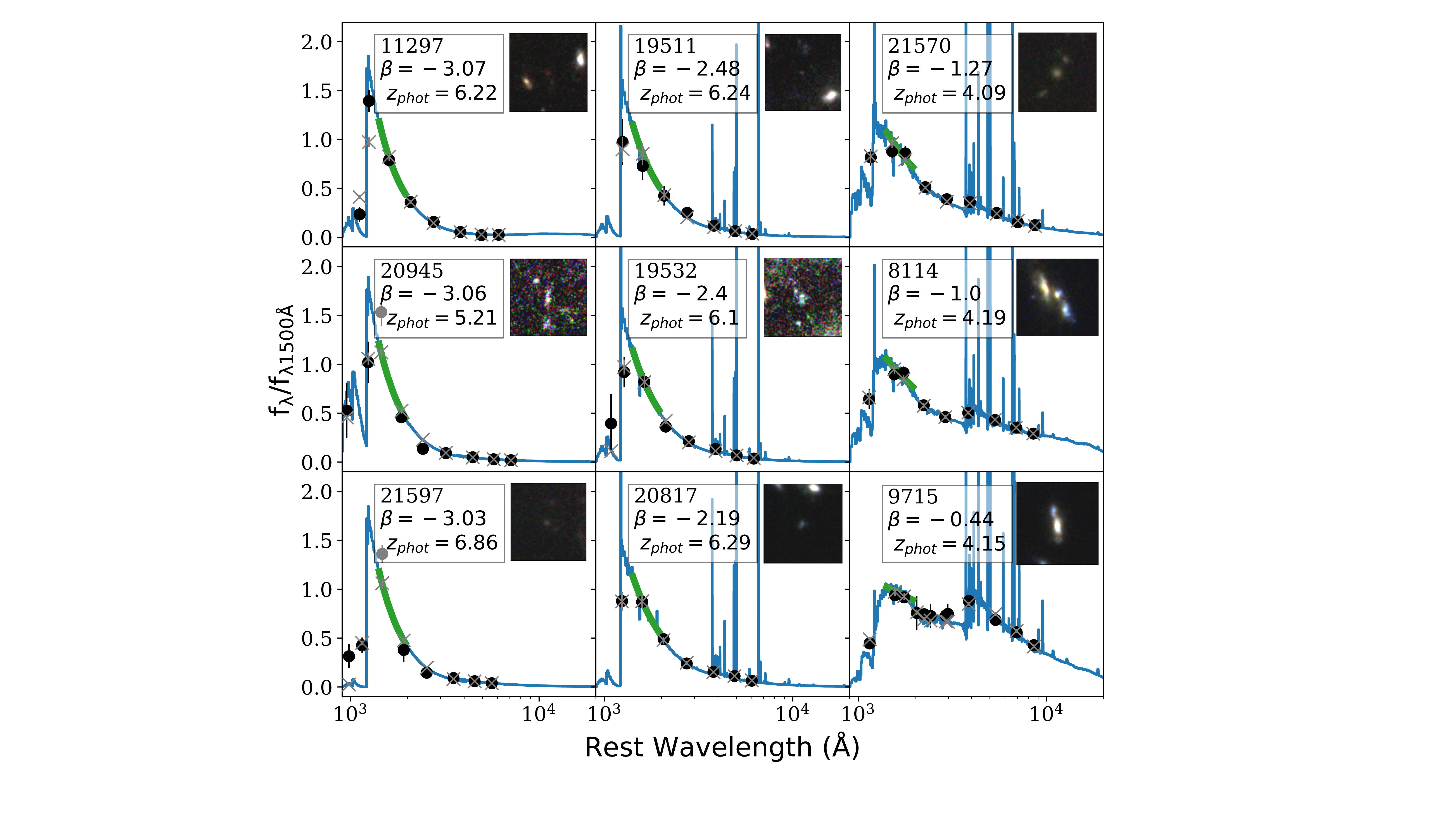}
\caption{
Here we show a representative sample of \eazy\ best-fit galaxy SEDs used in our analysis. 
Best-fit \eazy\ SED photometry and the real observed photometry with their associated errors are shown by crosses and circles, respectively. 
We show the fit to the UV slope in green.
The montages show rest-frame $g, r, i$ color images of the galaxies (see \citet{Jacobs2023a} for details). 
The text in each panel gives the galaxy ID as in \citep{Paris2023a} catalogue, the UV slope exponent, and the photometric redshift. 
\label{fig:seds}}
\end{figure*}

We present the \eazy\ derived rest-frame $u_s-g_s$ vs $g_s-i_s$ colors for our sample in Figure \ref{fig:colors}.
This color space has been shown to be effective in distinguishing between quiescent and star-forming galaxies \citep{Antwi-Danso2022a} and is an improvement over the rest-frame $U-V$ vs $V-J$ colors used at lower redshifts \citep{Williams2009,Schreiber2018b}. 
Based on our photometric redshifts, most of the galaxies in our sample fall in the blue star-forming region. 
There is one galaxy that falls at the edge between star-forming and quiescent region. 
Galaxies that have shallower UV slopes in general show redder  $u_s-g_s$ and $g_s-i_s$ colors.
We further compare the differences in $u_s-g_s$ and $g_s-i_s$ colors of galaxies selected by our NIRCam/$F444W$ selection and 3DHST $WFC3/F160W<26.4$ selection. 
We find that there is no significant difference between the color distributions of these two samples. 
This confirms that our selection is simply targeting fainter galaxies with similar underlying distribution in $u_s-g_s$ and $g_s-i_s$ colors. i.e. we are not finding a hidden population of red-dusty sources with the rest-frame $F444W$ selection. 
We note that at our highest redshift bin, the  $i_s$ band falls out of our $F444W$ filter. 
Therefore, the rest-frame color here is not directly constrained by the observed photometry.
In Figure \ref{fig:seds} we show \eazy\ best-fit SEDs of a representative sample of galaxies presented in our analysis.

We compute stellar masses for our sample using {\tt fast++} \citep{Schreiber2018} at the \eazy\ best-fit redshift. 
We use \citet{Bruzual2003} stellar population models with a \citet{Chabrier2003} IMF, a truncated SFH with a constant and an exponentially declining SFH component, and a \citet{Calzetti2000} dust law. 
We further derive stellar masses based on non-parametric SFHs \citep{Leja2019a} using {\tt Prospector} \citep{Johnson2021a} SED fitting similar to \citet{Leja2019b} analysis.  
The stellar masses derived by {\tt Prospector} includes emission line contribution to observed photometry. We find that the stellar masses between {\tt fast++} and {\tt Prospector} agree within the expected $\sim0.3$ dex uncertainty \citep{Conroy2013,Leja2019b}.

In Figure \ref{fig:beta_relationships}, we show the relationship of the UV continuum slope with redshift, UV magnitude, and stellar mass. 
In Table \ref{tab:binned_params} we tabulate the binned values observed for these three relationships. 
While we provide a qualitative description of our relationships, we refrain from quantifying them until after we conduct a through analysis of inherent biases associated with UV slope measurements at these redshifts \citep{Finkelstein2012a,Dunlop2013a,Bouwens2014a} in future work.

The UV slope of galaxies does not show any statistically significant correlation (parameterized by the Spearman correlation coefficient \citep{Spearman1904}) with redshift.
However, we find the bluest galaxies in our sample at higher redshifts and reaches a clear minimum value for $\beta\sim-3.1$. 
This may suggest a limitation on the input SED templates not being able to produce UV slopes blueward of $\lesssim-3.1$.

The UV magnitudes of our sample peaks at $M_{UV}\sim-19$. 
We observe no statistically significant correlation between UV slope and UV magnitude.  
Except at the brightest magnitudes, it is clear that some galaxies do reach a clear minimum for UV slopes.

We observe a statistically significant correlation between the UV slope and stellar mass. 
In general, galaxies with lower stellar masses tend to prefer bluer UV slopes. 
Given the buildup of stellar mass and dust is correlated, we expect higher stellar mass galaxies to have shallower UV slopes suggesting a higher amount of dust obscuration. 
We find no correlation between $u_s-g_s$ and $g_s-i_s$ colors with stellar mass in any of the redshift bins. This points to a lack of age/color correlation with stellar mass for our sample.  
Individual galaxies with stellar masses $\log_{10}(M_*/M_\odot)\lesssim 9$ show evidence of reaching the lower limit for UV magnitude allowed by the templates. 
We revisit this in Section \ref{sec:dust}.

\begin{deluxetable}{crrr}
\tabletypesize{\scriptsize}
\tablecaption{ Binned values for our galaxies. \label{tab:binned_params}}
\tablecolumns{4}
\tablewidth{0pt} 
\tablehead{ \colhead{Bin range}  & \colhead{N$_{\mathrm{sources}}$}     &  \colhead{Median $\beta$ }  &  \colhead{$\sigma_\beta$ }}
\startdata
$4.00<z<4.75$ & 89      &  -2.08    &  0.22  \\
$4.75<z<5.50$ & 140     &  -2.11    &  0.17  \\
$5.50<z<6.25$ & 91     &  -2.43    &  0.20  \\
$6.25<z<7.0$  & 81     &  -2.30    &  0.26  \\
\hline
$-23.52<M_{UV}<-20.07$  & 100    &  -2.07   &  0.22 \\
$-20.07<M_{UV}<-19.22$  & 100    &  -2.21   &  0.29 \\
$-19.22<M_{UV}<-18.66$  & 100    &  -2.27   &  0.18 \\
$-18.66<M_{UV}<-17.11$  & 101    &  -2.27   &  0.15 \\
\hline
$7.35<log_{10}(M_*/M_\odot)< 8.20 $ & 99     &  -2.48  &  0.17 \\
$8.20<log_{10}(M_*/M_\odot)< 8.63 $ & 101    &  -2.29  &  0.13 \\
$8.63<log_{10}(M_*/M_\odot)< 9.08 $ & 98     &  -2.11  &  0.15 \\
$9.08<log_{10}(M_*/M_\odot)< 10.36$ & 103    &  -1.86  &  0.24 \\
\enddata
\tablecomments{ Values computed for the binned samples as shown by Figure \ref{fig:beta_relationships}.}
\end{deluxetable}

\section{The nature of the UV slopes in the early Universe} \label{sec:dust}

The UV slope of a galaxy is constrained by the observed photometric bands and thus can be considered as a direct observable. 
Traditionally, the UV slope has been a useful measurement of the dust content in galaxies where dust emission in the infra-red cannot be easily constrained \citep[e.g][]{Reddy2006a}. 
At a fixed choice of an attenuation curve, the amount of reddening measures the amount of dust between the young blue O and B type stars in a galaxy and its close proximity \citep[e.g.][]{Calzetti1994,Meurer1999}. 
Shallower the UV slope (i.e. more positive $\beta$ values), higher the amount of dust obscuration that is expected in galaxies.  
However, degeneracies between galaxy age, mass, and luminosity could add extra uncertainty in deriving the amount of dust with the UV slope \citep[e.g][]{Reddy2010a,Casey2014a,Bouwens2016c}. 
While far-infrared (FIR) detections could help alleviate this tension, the infra-red excess of galaxies at $z>4$ is challenging to interpret due to the general reliance of a single ALMA detection to derive the total FIR luminosity.  
Additionally, constraining analysis to ones only with FIR detections biases analysis towards UV bright or heavily dust obscured galaxies \citep{Hodge2020}.

Our analysis is purely based on a $F444W$ magnitude selected sample of galaxies where we have confident S/N$>5$ detections in at least 4 photometric bands covering both rest-UV and optical wavelengths.
The $F444W$ band cover the rest-optical bands for our galaxies; thus can have significant contributions from emission lines and nebular continuum.
We have visually inspected all multi band photometric images,  SED fits, and the UV slope fits to every galaxy in the sample. 
Our stringent quality cuts on the {\eazy} best fit SEDs allow us to gain initial insights to the UV slopes of rest-optical selected galaxies at $4<z<7$.
However, we caution that the stringent quality cuts imposed on our magnitude selected sample could lead to incompleteness and selection effects. 
This needs to be modeled in terms of observed photometry of the NIRCam bands and is out of scope of this letter.

Our sample reaches UV magnitude of $M_{UV}\sim-18$, which is comparable to the \emph{Hubble} blank field data \citep[][]{Bouwens2022c}.  
A modest amount of lensing magnification is expected to be present in GLASS-JWST NIRCam parallel fields. 
In this initial set of papers we neglect the effect and will be revisited after the completion of the campaign. 
However, we stress that quantities such as the UV slope and colors are unaffected by magnification and therefore our conclusions are robust in that respect.

\begin{figure*}
\includegraphics[scale=0.6]{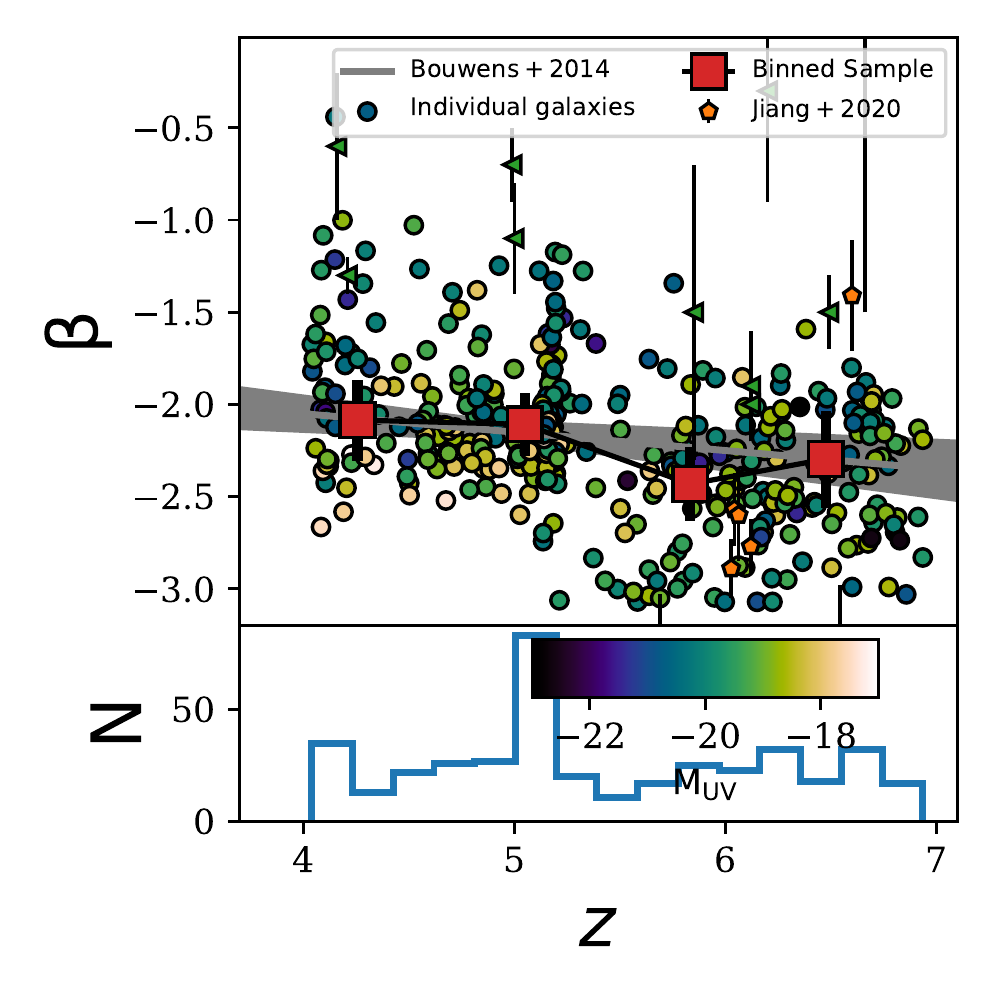}
\includegraphics[scale=0.6]{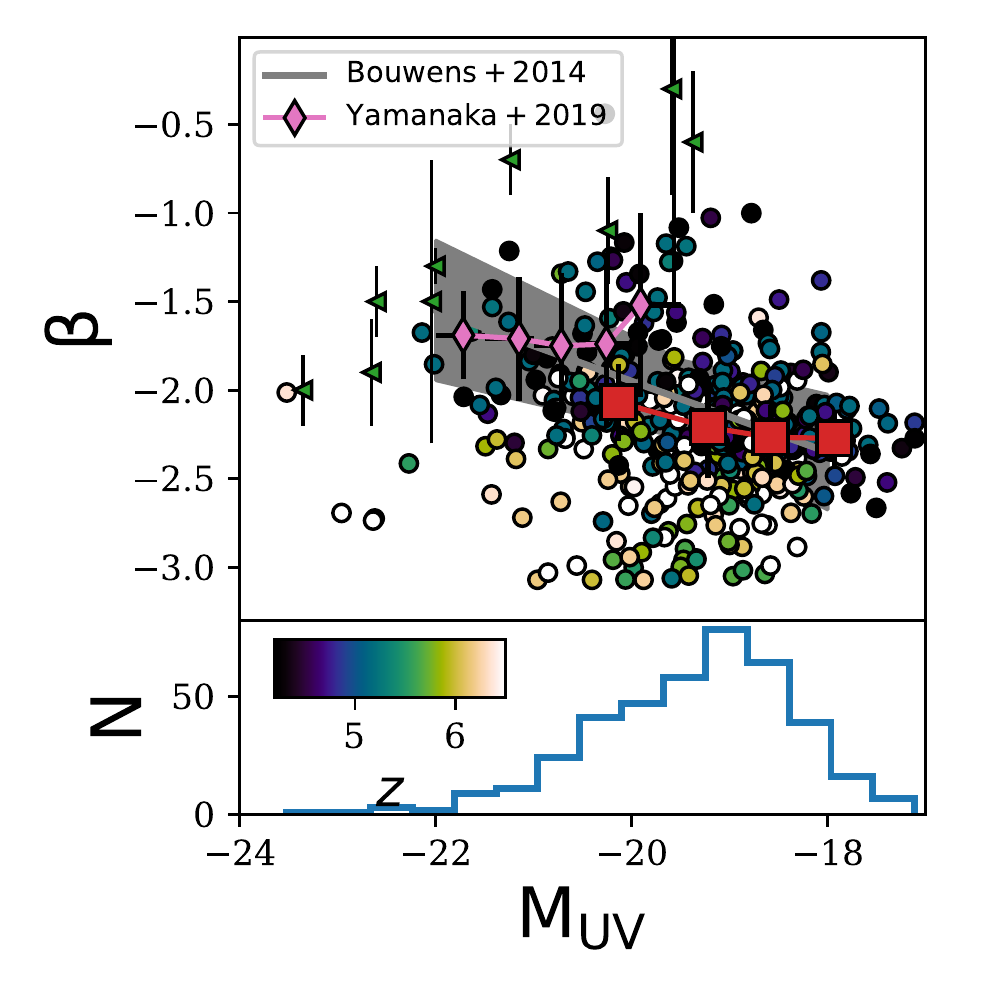}
\includegraphics[scale=0.6]{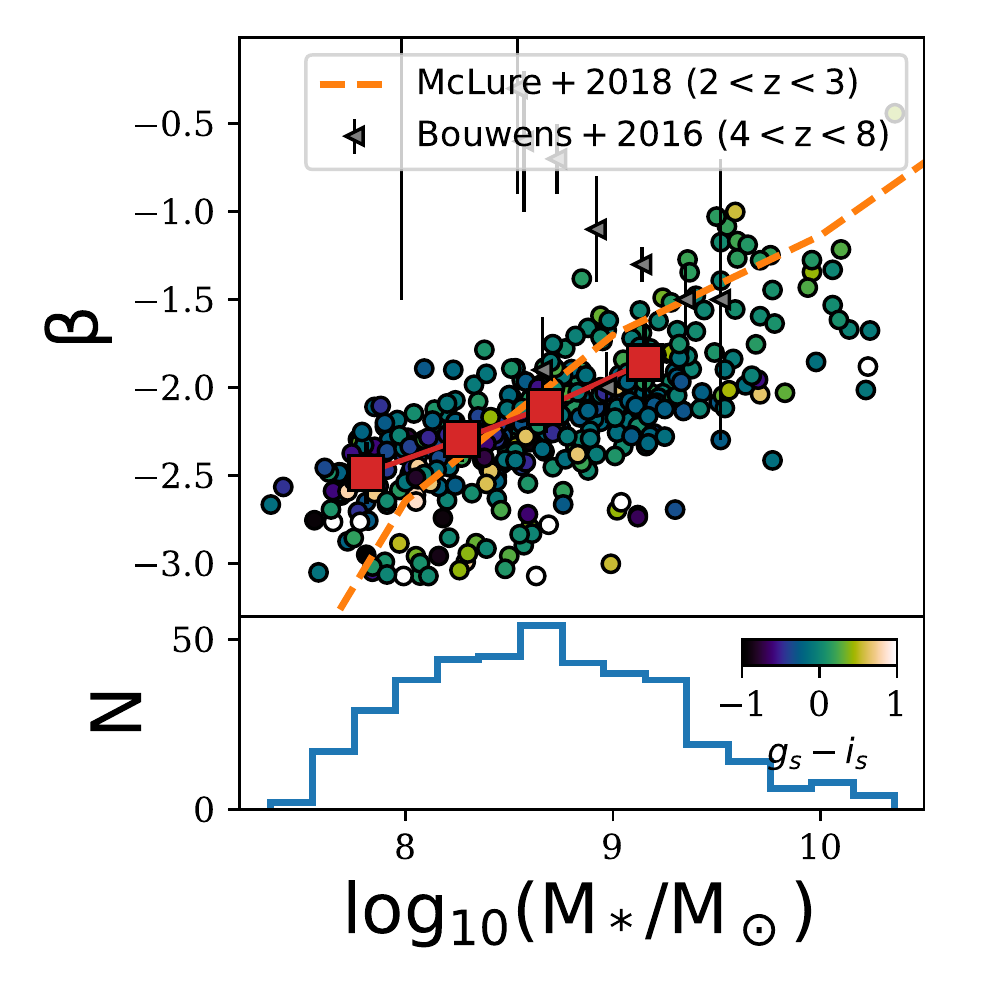}
\caption{The relationship of the UV slope $\beta$ with {\bf Left:} redshift (colour coded in terms of UV magnitude), {\bf Center:} rest-frame UV magnitude (colour coded in terms of redshift), and {\bf Right:} stellar mass (colour coded in terms of $g_s-i_s$ color). Galaxies are further binned in each of the three parameters and are shown in the respective panels. Values reported by \citet{Bouwens2016b,Yamanaka2019a,Jiang2020} as well as empirical relationships by \citet{Bouwens2014a,McLure2018b} are shown for comparison. 
We find no statistically significant evolution for $\beta$ with redshift or UV magnitude, however, we find that $\beta$ show a dependence on stellar mass. 
A fraction of galaxies reach $\beta\lesssim-3.1$ suggesting that SED templates with  $\beta<-3.1$ are required to accurately predict the rest-UV properties of some galaxies in this epoch. 
\label{fig:beta_relationships}}
\end{figure*}

\citet{Bouwens2014a} derived an empirical relationship between UV slope with redshift ($z\sim4-7$) based on deep \emph{HST} data reaching to $M_{UV}=-16.7$. 
Within our detection levels,  we do not observe a clear redshift or UV magnitude evolution of the UV slope for the GLASS-JWST data (Figure \ref{fig:beta_relationships}). 
\citet{Bouwens2016b} sample is based on infrared detected Lyman-break galaxies (LBGs), thus it is reasonable that these galaxies observe significantly shallower UV slopes compared to our sample.   
$z<3$ galaxies \citep{McLure2018b} also show shallower UV slopes compared to our sample, specially at $\mathrm{log_{10}(M_*/M_\odot)>9.0}$.

In Figure \ref{fig:beta_relationships} we also show the UV slopes of the luminous \Lalpha\ emitters observed by \citet{Jiang2020}.
Our sample also reach the extreme blue slopes observed in the \citet{Jiang2020} sample. 
\citet{Jiang2020} finds that the bluer galaxies in their sample with $\beta\sim-2.7$ are challenging to be reproduced with stellar population models. With the recent \citet{Larson2022a} templates used in our analysis, we are able to achieve blue UV slopes of $\sim-3.1$.

As mentioned in Section \ref{sec:UV_slope}, our sample at lower masses and higher redshifts  reach the bluer $\beta \lesssim-3.1$ limit that is capable to be fit even by the newest generation of \eazy\ models.  
This  is evident due to the horizontal limit that is observed at the bluer $\beta$ values in the individual measurements.  
Even though the SEDs of the very blue slopes are well constrained over multiple filters, it is possible for photometric redshift uncertainties to scatter redder galaxies to have bluer UV slopes. 
We perform bootstrap resampling of UV slope measurements to constrain the uncertainty associated with the $\beta$ measurements. 
For each galaxy, we recompute the SED fit 100 times within $1-\sigma$ photometric redshift uncertainty bounds defined by the \eazy\ $P(z)$ distributions. The $1-\sigma$ scatter of the measured $\beta$ values are taken as its error, $\Delta \beta$.
The median  $\Delta \beta$ for our sample is $0.034\pm0.11$. For galaxies with $\beta<-3.0$, $\Delta \beta =0.004\pm0.03$. 
Therefore, it is unlikely that the bluest UV slopes in our sample is a result of the redder galaxies being scattered to blue.    

We further test {\tt PEGASE} \citep{Fioc2019} and FSPS \citep{Conroy2010b} model templates with \eazy\ but are unable to resolve the limitation imposed by the templates to the observed UV slopes.
It is likely that a combination of Population III templates \citep[as investigated by][]{Bouwens2010b} or models considering UV continuum leakage/AGN effects as discussed by \citet{Jiang2020} could be required to reach bluer UV slopes than what is allowed by the templates used in our analysis. 
As the goal of this paper series is to demonstrate the capabilities of JWST data in the context of ERS, we defer a complete treatment of the analysis of UV slopes to future work.

In terms of UV magnitude, our galaxies span a large range in $M_{UV}\sim-23$ to $\sim-18$. 
Our  sample does not show a statically significant correlation between UV magnitude and the UV slope.
However, we refrain from interpreting this further due to selection effects that could have arisen from our bluer filters. 
While magnitude depths are quite constant across the bands and reach $\gtrsim30$th magnitude, $F090W$ being the bluest band is slightly shallower. 

Previous  works have found conflicting results on the UV magnitude dependence on $\beta$ \citep[e.g.][]{Finkelstein2012a,Bouwens2014a,Yamanaka2019a,Bhatawdekar2021a}, especially at $z>6$. 
Our binned values are largely in agreement with the \citet{Bouwens2014a} relation at brighter UV magnitudes. 
The $z\sim4$ LBGs in \citet{Yamanaka2019a} sample shows redder UV slopes compared to our results. 
The stellar population analysis of LBGs show that they have a young star-forming stellar population with high amounts of dust.
From the $ugi$ color analysis in Figure \ref{fig:colors}, we found that GLASS-JWST data are preferentially biased toward blue star-forming systems. Thus, we can rule out significant dust obscuration in our galaxies, and the distinction with \citet{Yamanaka2019a} is as expected.

In order to determine the dust buildup in the early Universe, the UV slope evolution should be linked with stellar mass and age. 
As validated by the rest-frame color distributions, we expect the majority of galaxies to be blue, young star-forming systems at these redshifts. 
Simulations by \citet{Popping2017a} show that stellar age also plays a significant role in determining the UV slope. 
Given sources with redder UV slopes show a bias toward higher masses (Figure \ref{fig:beta_relationships}) we cannot rule out effects from age to our analysis. 


Most of our galaxies have blue UV slopes. 
The expected median attenuation following the \citet{Meurer1999} relation is $Av_{\beta}=0.01\pm0.33$.
Therefore, based on our GLASS-JWST NIRCam $F444W$ selected sample, we expect galaxies at $4<z<7$ to be relatively dust-free blue star-forming systems. 
Within this window the Universe was only $\sim0.7-1.5$ Gyrs old, thus we expect supernovae to be the primary driver of dust buildup in these early systems \citep{Hodge2020}. 
When compared to $z\sim2$ $K$-band selected samples \citep[e.g][]{Shivaei2018,Nanayakkara2020}, the UV slopes observed by our survey are significantly bluer.   
At later times, we expect AGB stars to be the dominant dust production mechanisms leading the general population of galaxies to be more dust rich.

We refrain from linking our UV slopes to expected  infra-red excess (IRX)  at $z>4$. 
As shown by \citet{Bouwens2016b}, the increase of dust temperature with redshift \citep{Bethermin2015a,Schreiber2018c} and correlations with the assumed dust attenuation curve could add biases to the UV slope-IRX relationship.
If galaxies at early times show harder ionizing spectra with high-energy emission lines \citep[e.g][]{Mainali2018,Nanayakkara2019}, the relationship between dust production and relationship becomes important. 
Furthermore, any evolution of the IMF with redshift \citep{Nanayakkara2017,Sneppen2022a} would add further complications to interpreting the UV slope-dust evolution with redshift.

\section{Conclusions and Future Work} \label{sec:conclusions}

We have presented a first view from JWST on the UV slope evolution of NIRCam $F444W$ selected galaxies at $z\sim4-7$ from our GLASS-JWST program. 
\begin{itemize}
\item We find that a majority of our galaxies ($>99\%$) are  blue star-forming systems with steep UV slopes. 
\item We find no statistically significant evolution of the UV slopes with redshift or UV magnitude. 
\item We find a statistically significant positive correlation of UV slope with stellar mass. 
\item We find that galaxies with faint UV magnitudes and low stellar masses have bluer UV slopes compared to their UV bright high mass counterparts. This suggest that faint UV systems observed by JWST are blue, low SFR systems.
\item We find that some individual measurements of UV slopes hit the bluer limit $\beta \lesssim -3.1$ imposed by the SED templates used in our analysis. 
\end{itemize}

Our observations point to a presence of very blue galaxies in the $z\sim4-7$ Universe. 
Stellar population models derived using BPASS \citep{Eldridge2017} and FSPS \citep{Conroy2010b} are able to obtain the UV slope properties of most of our galaxies. 
Very blue slopes $\beta\sim-3.1$ can be produced by stellar only templates (no nebular continuum contribution) and/or templates with Lyman continuum leakage \citep{Topping2022a}.
Additionally models that include Population III dust free stars \citep[e.g][]{Bouwens2010b} can also produce very blue UV slopes. 
Within the context of our work we find that most of the UV slopes of our galaxies can be explained by current stellar population models without the need for any strong exotic effects.

Here we have found that rest-optical selected galaxies at $4<z<7$ are preferentially star-forming galaxies with steep UV slopes. 
Future work should focus on sample selection effects to determine the real number density of extreme blue systems in the early Universe. 
By combining deep NIRCam observations between different ERS and JWST Cycle 1 treasury programs, a large representative sample of galaxies at these redshifts could be constructed to analyze the evolution of the UV slope with redshift and UV magnitude. 
Future spectroscopic observations would be crucial to determine what causes the very blue UV slopes of these galaxies. i.e. high Lyman continuum leakage, extreme low metallicities. 
Obtaining constraints to the infrared emission and dust temperatures using deep JWST/MIRI and ALMA observations \citep{Schreiber2018c} planned on JWST JWST Early Release Science (ERS)/Treasury fields and inter-stellar-medium (ISM) conditions from JWST/NIRSpec observations would add important diagnostic power to determine how the evolution of  mass and dust was modulated in the early Universe.

\facility{JWST: (NIRCam)}

\begin{acknowledgments}
This work is based on observations made with the NASA/ESA/CSA James Webb Space Telescope. The data were obtained from the Mikulski Archive for Space Telescopes at the Space Telescope Science Institute, which is operated by the Association of Universities for Research in Astronomy, Inc., under NASA contract NAS 5-03127 for JWST. These observations are associated with program JWST-ERS-1342. We acknowledge financial support from NASA through grant JWST-ERS-1324.
T.N., K. G., and C.J. acknowledge support from Australian Research Council Laureate Fellowship FL180100060.  MB acknowledges support from the Slovenian national research agency ARRS through grant N1-0238. CM acknowledges support by the VILLUM FONDEN under grant 37459. The Cosmic Dawn Center (DAWN) is funded by the Danish National Research Foundation under grant DNRF140.
This project made use of {\tt astropy} \citep{Astropy2018}, {\tt matplotlib} \citep{Hunter2007}, and {\tt pandas}  \citep{Pandas2020}.
\end{acknowledgments}

\bibliographystyle{aasjournal}
\bibliography{bibliography.bib}

\end{document}